\documentclass{article}
\usepackage[utf8]{inputenc}
\usepackage[english]{babel}
\usepackage[margin=0.8in]{geometry}
\usepackage{graphicx}
\usepackage{amsmath}
\usepackage{amssymb}

\begin{document}

\title{Lorentz violation in a uniform Newtonian gravitational field}

\author{Y. Bonder\footnote{email: ybonder@indiana.edu}\\Physics Department, Indiana University, Bloomington, IN 47405, USA}

\date{(Dated: IUHET 582, September 2013)}

\maketitle
\begin{abstract}
Lorentz invariance is one of the fundamental principles of physics, and, as such, it must be experimentally tested. The purpose of this work is to obtain, within the Standard-Model Extension, the dynamics of a Lorentz-violating spinor in a uniform Newtonian gravitational field. This is achieved by treating the spinor as a test particle and introducing the gravitational field through a uniformly accelerated observer. The nonrelativistic Hamiltonian is obtained, and some experimental consequences are discussed. One unexpected outcome of this work is that the gravitational field helps disentangle bounds on coefficients for Lorentz violation.
\end{abstract}

\section{Introduction}

Lorentz invariance is part of the foundation of modern physics rendering its empirical validation extremely important. The Standard-Model Extension (SME) is a framework within effective field theory \cite{KosteleckyPotting} that incorporates all possible Lorentz- (and CPT \cite{Greenberg}) violating extensions to General Relativity and the Standard Model \cite{ColladayKostelecky,Kostelecky04}. A Lorentz-violating term in the SME is formed by a Lorenz-violating operator, built from conventional fields, contracted in a coordinate independent way with a controlling coefficient. The SME has been widely used for tests of Lorentz invariance \cite{Review Liberati}; a list of experiments and constraints can be found in Ref.~\cite{DataTables}.

The SME coefficients may be ``explicit,'' namely, objects present in the theory without being dynamical, or may be ``spontaneous'' and arise when certain fields attain its vacuum expectation values. In the latter case, the coefficients have fluctuations about its vacuum expectation value that may have physical implications \cite{NG}. As proven by Kosteleck\'y \cite{Kostelecky04}, explicit breaking is incompatible with the Bianchi identity, which, in turn, lies at the core of Einstein's equations. Therefore, Lorentz violation in the presence of gravity requires spontaneous coefficients.

One of the original motivations for introducing gravity in the context of Lorentz-violating spinors is that some SME coefficients can only be empirically accessed through couplings with gravity \cite{KosteleckyTasson,KosteleckyTasson09}. In this case it is necessary to solve Einstein's equations and consider dynamical mechanisms for Lorentz violation, making the calculations extremely challenging. Kosteleck\'y and Tasson \cite{KosteleckyTasson} were able to study Lorentz-violating spinors in the presence of a general gravitational field under the linearized metric approximation, leading to the first bounds on some coefficients. However, in Ref.~\cite{KosteleckyTasson} the SME coefficients associated with spin are disregarded.

The goal of the current work is to incorporate gravity to the analysis of Lorentz-violating spinors without neglecting the SME coefficients associated with spin. There are two reasons that render this work particularly interesting. First, spin tests of Lorentz invariance are remarkably sensitive \cite{boundsb}, and it is conceivable that, through gravity couplings, the bounds set with these experiments could be translated into more stringent bounds on other SME coefficients. Second, the effects of quantum gravity could become manifest when gravity and an inherit quantum property of matter---like spin---are simultaneously present.

To include spin with the techniques developed in Ref.~\cite{KosteleckyTasson} seems daunting. Thus, the challenge is to find approximations which properly describe the experimental conditions and, at the same time, allow one to calculate the Hamiltonian for Lorentz-violating spinors in the presence of gravity. A method achieving this goal is presented in Sec. \ref{formalism}. Since most experiments testing Lorentz invariance are properly described in the nonrelativistic regime, the nonrelativistic Hamiltonian is also obtained. Furthermore, an analysis of the physical content and several consistency checks are presented in Sec. \ref{formalism}. Section \ref{ExpConseq} contains a brief overview of the experimental consequences of the resulting Hamiltonian, including a discussion on the possibility of using gravity to disentangle bounds on linear combinations of SME coefficients. The conclusions are presented in Sec. \ref{Conclusions}. Finally, the notation and conventions used throughout the paper are described in Appendix \ref{Notation}, and an analysis of field redefinitions is given in Appendix \ref{FieldRedef}.

\section{Formalism}\label{formalism}

\subsection{Starting point}

This section is devoted to deriving the Hamiltonian describing the evolution of a Lorentz-violating spinor in the gravitational environment of a laboratory. Note that, in contrast with other research programs \cite{BaileyKostelecky}, in this work the spacetime metric is not assumed to be Lorentz-violating. In addition, the mechanisms that could generate SME coefficients, which have been studied in contexts such as string theory \cite{LV strings}, noncommutative spacetime \cite{LV noncommutativity}, loop quantum gravity \cite{LV LQG}, nonminimal gravitational couplings \cite{QGP}, and through spontaneous symmetry breaking of vector and tensor fields \cite{LV SSB}, are not the concern of the present work. Here, Lorentz violation occurs in the spinor field and its sources are not considered.

Under the test-particle approximation, the contribution of the spinors to the spacetime curvature is disregarded and their energy-momentum tensor does not enter into Einstein's equations. Being a good approximation for most laboratory experiments, this approximation is assumed, and the theorem forcing the SME coefficients to arise spontaneously can be circumvented. In other words, under the test-particle approximation, the fluctuations of the SME coefficients can be consistently neglected, making the calculations manageable.

Moreover, in this work the attention is focused on a free Dirac spinor $\psi$, and only Lorentz-violating terms of renormalizable dimensions are considered. This subset of the SME is called the minimal matter sector, and the spinor action in a general curved spacetime background is given by \cite{Kostelecky04}
\begin{subequations}\label{action}
\begin{equation}
S=\int d^4 x e \Big[\frac{i}{2}e^\mu_a \bar{\psi}\Gamma^a \nabla_\mu 
\psi-\frac{i}{2}e^\mu_a (\nabla_\mu\bar{\psi})\Gamma^a \psi-\bar{\psi}M\psi\Big],
\end{equation}
where $\Gamma^a$ and $M$ are defined as
\begin{eqnarray}
\Gamma^a&=&\gamma^a-c_{\mu\nu}\eta^{ac}e^\mu_b e^\nu_c \gamma^b-d_{\mu\nu}\eta^{ac}e^\mu_b e^\nu_c \gamma_5 \gamma^b - e_\mu 
\eta^{ac}e^\mu_c-i f_\mu \eta^{ac}e^\mu_c \gamma_5-\frac{1}{2}g_{\mu\nu\rho}\eta^{ab} e^\mu_c e^\nu_d e^\rho_b \sigma^{cd},\\
M&=&m+i m_5 \gamma_5+a_\mu e^\mu_a \gamma^a+b_\mu e^\mu_a \gamma_5 \gamma^a+\frac{1}{2}H_{\mu\nu} e^\mu_a e^\nu_b \sigma^{ab}.
\end{eqnarray}
\end{subequations}
The explicit form of the covariant derivatives $\nabla_\mu$ acting on $\psi$ is presented in Appendix \ref{Notation}. In this case, the SME coefficients are $a_\mu$, $b_\mu$, $c_{\mu\nu}$, $d_{\mu\nu}$, $e_\mu$, $f_\mu$, $g_{\mu\nu\rho}=-g_{\nu\mu\rho}$, and $H_{\mu\nu}=-H_{\nu\mu}$, which are assumed to have small components in any experimentally relevant coordinate system \cite{Ralf}. The chiral mass $m_5$ is only included for the purpose of generality; for simplicity, it is also considered small.

The equation of motion for the spinor obtained from the action (\ref{action}) is 
\begin{eqnarray}
0&=&ie^\mu_a \Gamma^a \partial_\mu \psi+\frac{i}{2} e^\mu_a \omega_{\mu cd} 
\left(\eta^{ac}\Gamma^d +\frac{i}{4}\{\Gamma^a,\sigma^{cd}\}\right)\psi+\frac{i}{2}e^\mu_a (\partial_\mu \Gamma^a) \psi-M\psi,\label{eom}
\end{eqnarray}
with $\omega_{\mu ab}$ being the spin connection (see Appendix \ref{Notation}). The goal is to find a Hamiltonian for $\psi$ defined as the time-evolution operator $i\partial_0$. To do so, it is necessary to restrict the analysis to a static (and torsion-free) spacetime where there is no contribution to the spinor evolution from spacetime. Before doing so, it is important to remark that, from this point on, $\psi$ is regarded as a spinorial wave function in the context of one-particle quantum mechanics and not as a quantum field. 

\subsubsection{General static spacetime}

In any static spacetime the tetrad can be chosen as \cite{Wald}
\begin{equation}
e^\mu_0=\delta^\mu_0 e^0_0(x^k) , \qquad e^\mu_j=\delta_i^\mu e^i_j(x^k) ,
\end{equation}
where $e^0_0\neq0$. Therefore, the equation of motion (\ref{eom}) can be brought to the form
\begin{equation}
0=ie^0_0 \Gamma^0 \partial_0 \psi+ie^i_j \Gamma^j \partial_i \psi+\frac{i}{2} e^\mu_a \omega_{\mu cd} 
\left(\eta^{ac}\Gamma^d +\frac{i}{4}\{\Gamma^a,\sigma^{cd}\}\right)\psi+\frac{i}{2}e^\mu_a (\partial_\mu \Gamma^a) \psi-M\psi.
\end{equation}
Note that, if $e^0_0 \Gamma^0$ is inverted, a Hamiltonian can be read from this equation. 

The inner product associated with such a Hamiltonian can be obtained by inspecting the corresponding continuity equation \cite{KosteleckyTasson,Parker}, and, in general, it is not the standard inner product of nonrelativistic quantum mechanics,
\begin{equation}\label{inner product}
 \left<\psi_1,\psi_2 \right> = \int_\Sigma dv \psi_1^\dagger \psi_2,
\end{equation}
where $\psi_1$ and $\psi_2$ are any (square-integrable) spinors and $\Sigma$ is a Cauchy surface for which the natural volume element is $dv$. Since the goal is to find a Schr\"odinger-like equation, it is convenient to use a method yielding to the inner product (\ref{inner product}). It has been shown \cite{KosteleckyTasson} that this can be achieved with a field redefinition $ \psi = W \chi$, where $W$ is chosen so that the factor of $i\partial_0 \chi$ becomes $e^0_0 \gamma^0$, which, in addition, can be easily inverted.

Observe that, in contrast to the case where gravity is disregarded \cite{RalfFR}, here it is difficult to find a field redefinition valid to all orders in the SME coefficients. However, it is possible to verify that $W=(3-\gamma^0\Gamma^0)/2= W^\dagger$ works to first order in $m_5$ and the SME coefficients, which is an approximation used for the rest of this paper. The Hamiltonian obtained with this method takes the form
\begin{subequations}\label{Hamiltonian static Hermitian}
\begin{equation}
H = -i\frac{e^i_j}{e^0_0} \gamma^0\widetilde{\Gamma}^j \partial_i + \frac{1}{e^0_0} \gamma^0\widetilde{M},
\end{equation}
where
\begin{eqnarray}
\bar{W}&=& \gamma^0 W^\dagger \gamma^0,\\
\widetilde{\Gamma}^a &=& \bar{W} \Gamma^a W,\\
\widetilde{M} &=& \bar{W} M W - i e^\mu_a \bar{W}\gamma^a (\partial_\mu W)-\frac{i}{2}e^\mu_a \bar{W} (\partial_\mu \Gamma^a)W -\frac{i}{2}e^\mu_a\omega_{\mu cd}\bar{W} \left(\eta^{ac}\Gamma^d +\frac{i}{4}\{\Gamma^a,\sigma^{cd}\}\right)W.
\end{eqnarray}
\end{subequations}

\subsubsection{Uniform Newtonian gravitational field}

The background spacetime has to be chosen to properly characterize the gravitational environment in laboratory experiments. This can be achieved by a uniform Newtonian field, which, in turn, is described by a flat spacetime as seen by a uniformly accelerated observer provided that the observer's acceleration is identified with the gravitational acceleration. The corresponding metric can be written \cite{Hehl90,Bonder13} as
\begin{equation}\label{metric}
 ds^2=-(1+\Phi)^2dt^2+dx^i dx_i,
\end{equation}
where $\Phi$ is the uniform Newtonian potential satisfying $\partial_0 \Phi=0$ and $\partial_i \partial_j \Phi =0$. That this metric accurately describes the gravitational environment in laboratory experiments can be justified from the fact that the Schwarzschild metric in Fermi-like coordinates associated with a fixed observer at the Earth's surface, takes the form of Eq.~(\ref{metric}) plus the curvature tensor contracted with two powers of a suitably defined distance \cite{Fermi coord}. These additional terms generate tidal effects and can be neglected when the size of the experiment is much smaller than the spacetime-curvature radius, as in most laboratory experiments. In fact, it has been explicitly shown that the metric (\ref{metric}) properly represents gravity in certain laboratory experiments (see Refs.~\cite{Daniel,Greenberger}).

Moreover, given that the spacetime under consideration is flat, it can be assumed that the covariant derivatives of the SME coefficients vanish globally, which is the generalization of the condition that, in the absence of gravity, the partial derivatives of the coefficients with respect to a Minkowskian reference frame are zero.

For the metric (\ref{metric}) the tetrad can be chosen as
\begin{equation}
e_0^0 = (1+\Phi)^{-1}, \qquad e^i_j = \delta^i_j.
\end{equation}
In addition, it is possible to check that the nonvanishing Christoffel symbols are
\begin{eqnarray}\label{Christoffels}
 \Gamma^0_{0i}= \Gamma^0_{i0}=(1+\Phi)^{-1}(\partial_i \Phi),\qquad
 \Gamma^i_{00}=\eta^{ij} (1+\Phi)(\partial_j \Phi),
\end{eqnarray}
and thus
\begin{equation}
e^\mu_a {\omega_\mu}_{cd}= \frac{(\partial_i 
\Phi)}{1+\Phi}\delta^0_a\left(\delta^i_c \delta^0_d- \delta^0_c 
\delta^i_d \right).
\end{equation}
Taking this into the account, the Hamiltonian (\ref{Hamiltonian static Hermitian}) takes the form
\begin{subequations}\label{relativistic Hamiltonian}
\begin{equation}
H = -i(1+\Phi)\gamma^0\widetilde{\Gamma}^i \partial_i + (1+\Phi) \gamma^0\widetilde{M},
\end{equation}
where
\begin{eqnarray}
\widetilde{\Gamma}^i &=&\gamma^i+\Gamma^i+\frac{1}{2}\left[\gamma^0\gamma^i,\Gamma^0\right],\\
\widetilde{M}&=&m+M-\frac{m}{2}\{ \gamma^0, \Gamma^0\}-\frac{i}{2}\gamma^0 \gamma^i(\partial_i 
\Gamma^0)  - \frac{i}{2} (\partial_i\Gamma^i)- \frac{i}{2} \frac{(\partial_i \Phi)}{1+\Phi} \left( \gamma^i + \Gamma^i +\gamma^0\gamma^i\Gamma^0\right).
\end{eqnarray}
\end{subequations}
To first order in $m_5$ and the SME coefficients, the relativistic Hamiltonian for a Lorentz-violating spinor in a uniform Newtonian gravitational field is given in Eqs.~(\ref{relativistic Hamiltonian}). Bear in mind that, even though Lorentz invariance is associated with relativity, the goal of this paper is to set the framework to look for experimental evidence of Lorentz violation in the presence of a uniform Newtonian gravitational field in experiments where the particles have nonrelativistic velocities, and can thus be described by a Schr\"odinger equation. This is why the nonrelativistic limit is sought, which is done next.

\subsection{Nonrelativistic limit}

The standard procedure to calculate the nonrelativistic Hamiltonian when dealing with spinors in the context of relativistic quantum mechanics is through a series of unitary transformations \cite{FoldyWouthuysen}. The idea is to write the Hamiltonian as
\begin{equation}
 H= m\gamma^0+\mathcal{E} + \mathcal{O},
\end{equation}
where $\mathcal{E}$ and $\mathcal{O}$ are respectively called the even and odd parts of the Hamiltonian; $\mathcal{O}$ is defined as the part of the Hamiltonian that couples particle and antiparticle degrees of freedom. Thus, in the nonrelativistic limit, $\mathcal{O}$ is expected to vanish. In general it is extremely hard to find a unitary transformation that eliminates $\mathcal{O}$. The alternative is to perform a series of unitary transformations, called Foldy--Wouthuysen transformations, each of which removes from $\mathcal{O}$ the leading contribution in $\partial_i/m$. After three iterations, the Hamiltonian, containing terms with two or less spatial derivatives, takes the form
\begin{equation}
 H_{\rm{FW}}=m\gamma^0+\mathcal{E}+\frac{1}{2m}\gamma^0 
 \mathcal{O}^2-\frac{1}{8m^2}[\mathcal{O},[\mathcal{O},\mathcal{E}]] -\frac{i}{8m^2}[\mathcal{O},\partial_0\mathcal{O}].\label{H FW general}
\end{equation}
Note that this approximation properly describes nonrelativistic experiments since $\partial_i$ may act on either the wave function, in which case it may be regarded as the particle's momentum, or on the Newtonian potential, resulting in the gravitational acceleration. For nonrelativistic experiments, the momentum of the particle and the gravitational acceleration ($\sim 10^{-32}\ \rm{GeV}$) are small with respect to $m$, and thus the aforementioned truncation is justified.

To illustrate the method, the nonrelativistic Hamiltonian is calculated in the absence of Lorentz violations and when $m_5=0$. In this case, the relativistic Hamiltonian is
\begin{equation}\label{H LI}
 H_{\rm{LI}}=-i \gamma^0 \gamma^i (1+\Phi)\partial_i +m \gamma^0 (1+\Phi) - 
 \frac{i}{2} \gamma^0\gamma^i(\partial_i \Phi),
\end{equation}
where the subindex ${\rm LI}$ is a reminder that this quantity is associated with the Lorentz-invariant and $m_5=0$ case. To calculate the nonrelativistic Hamiltonian, the even and odd parts of the Hamiltonian must be identified. In this case, these parts are given by
\begin{eqnarray}
 \mathcal{E}_{\rm{LI}}&=& m\gamma^0\Phi,\\
 \mathcal{O}_{\rm{LI}}&=&-i \gamma^0 \gamma^i \left[(1+\Phi) \partial_i + 
 \frac{1}{2} (\partial_i \Phi) \right].
\end{eqnarray}
Inserting these expressions in Eq.~(\ref{H FW general}) leads to
\begin{eqnarray}
H_{\rm{FW,LI}}&=&m\gamma^0(1+\Phi)+\frac{\eta^{ij}}{2m}\gamma^0 
\Big[-(1-\Phi)(1+\Phi)^2\partial_i 
\partial_j+\frac{1}{2}(1+2\Phi)(\partial_i\Phi) (\partial_j\Phi) -(1-3\Phi) 
(1+\Phi)(\partial_i\Phi) \partial_j \Big]\nonumber\\
&&-\frac{i{\epsilon_k}^{ij}}{4m} \gamma^0 \Sigma^k 
(1-3\Phi)(1+\Phi)(\partial_i\Phi) \partial_j, \label{LI FW Hamiltonian}
\end{eqnarray}
which is the nonrelativistic Hamiltonian for a Lorentz-invariant spinor in a uniform gravitational field.

The next step is to include $m_5$ and the SME coefficients. For that purpose, $\mathcal{E}$ and $\mathcal{O}$ must be read from the Hamiltonian (\ref{relativistic Hamiltonian}). Then these parts are substituted into Eq.~(\ref{H FW general}), retaining only the leading contributions in $m_5$ and the SME coefficients. Since only linear terms in these quantities are sought, the calculations are done one coefficient at a time. Recall that the covariant derivatives of the SME coefficients vanish by assumption, which is used to calculate the partial derivatives of these coefficients. 

The contribution of $m_5$ to the nonrelativistic Hamiltonian is
\begin{eqnarray}
\Delta H_{\rm{FW}}&=&-\frac{ m_5}{4m}\Sigma^k ( 1 + \Phi )(1-3\Phi)(\partial_k 
\Phi),\label{contribution m5}
\end{eqnarray}
which must be added to the Lorentz-invariant nonrelativistic Hamiltonian (\ref{LI FW Hamiltonian}). The same calculation is repeated for all the SME coefficients. The correction to the nonrelativistic Hamiltonian coming from $a_\mu$ is
\begin{equation}
\Delta H_{\rm{FW}}= a_0 -\frac{ia_i\eta^{ij}}{m}\gamma^0 
(1+\Phi)^2\left(1-\Phi \right)\partial_j-\frac{ia_i \eta^{ij}}{2m}\gamma^0 
(1+\Phi)(1 - 3\Phi)(\partial_j\Phi)-\frac{a_i}{4m}{\epsilon_k}^{ij} \gamma^0\Sigma^{k}(1+\Phi) (1 - 3\Phi)(\partial_j\Phi).
\end{equation}
Similarly, the contribution of $b_\mu$ to the nonrelativistic Hamiltonian is
\begin{eqnarray}
\Delta H_{\rm{FW}}&=&-b_i(1+\Phi)\Sigma^i +i\frac{b_0}{2m}\gamma^0 \Sigma^k 
\left(2(1-\Phi^2)\partial_k+(1-3\Phi)(\partial_k\Phi)\right) -i\frac{3\epsilon^{ikl}b_i}{4m^2}(1+\Phi)^2(\partial_k\Phi) \partial_l\nonumber\\
&&+\frac{b_i}{2m^2}(\eta^{il}\Sigma^k-\eta^{kl}\Sigma^i)\Big[(1+\Phi)^3\partial_k 
\partial_l +(1+\Phi)(\partial_k\Phi)(\partial_l\Phi) +\frac{3}{2}(1+\Phi)^2((\partial_k\Phi) \partial_l+(\partial_l\Phi) \partial_k)\Big].
\end{eqnarray}

When calculating the effects of $c_{\mu\nu}$, it is possible to study the irreducible components, $c_{(\mu\nu)}$ and $c_{[\mu\nu]}$, separately. For $c_{(\mu\nu)}$ the additional piece for the nonrelativistic Hamiltonian is
\begin{eqnarray}
\Delta H_{\rm{FW}}&=&-mc_{00}\gamma^0(1+\Phi)^{-1} -\frac{\eta^{ij}}{2m}c_{00}\gamma^0 
\Big[-(1-\Phi)\partial_i \partial_j+\frac{1}{2}\frac{1+2\Phi}{(1+\Phi)^{2}}(\partial_i\Phi) (\partial_j\Phi)-\frac{1-3\Phi}{1+\Phi}(\partial_i\Phi) \partial_j \Big]\nonumber\\
&&+\frac{i{\epsilon_k}^{ij}}{4m} c_{00}\gamma^0 \Sigma^k \frac{1-3\Phi}{1+\Phi}(\partial_i\Phi) \partial_j +2i\eta^{ij}c_{(0i)}(1+\Phi)^{-1}\left[(1+\Phi) \partial_j+\frac{1}{2}(\partial_j \Phi)\right]\nonumber\\
&&+\frac{1}{2}{\epsilon_k}^{ij}c_{(0i)}\Sigma^k(1+\Phi)^{-1}(\partial_j \Phi)-i\frac{{\epsilon_m}^{ik}\eta^{jl}c_{(ij)}}{4m}\gamma^0\Sigma^m (1+\Phi)(1-3\Phi)\left[(\partial_k \Phi) \partial_l- (\partial_l \Phi) \partial_k \right]\nonumber\\
&&+\frac{\eta^{ik}\eta^{jl}c_{(ij)}}{2m} \gamma^0
\Big[2(1-\Phi)(1+\Phi)^2 \partial_k \partial_l +(1-3\Phi)(1+\Phi)\left((\partial_k \Phi) \partial_l + (\partial_l \Phi) \partial_k\right)-(1+2\Phi)(\partial_k \Phi)(\partial_l \Phi) \Big],\nonumber\\
\end{eqnarray}
and the contribution of $c_{[\mu\nu]}$ is
\begin{equation}
\Delta H_{\rm{FW}}=-\frac{1}{2}{\epsilon_k}^{ij}c_{[0i]}\Sigma^k(1+\Phi)^{-1}(\partial_j \Phi) -i\frac{{\epsilon_m}^{ik}\eta^{jl}c_{[ij]}}{4m}\gamma^0\Sigma^m (1+\Phi)(1-3\Phi)\left[(\partial_k \Phi) \partial_l- (\partial_l \Phi) \partial_k \right].
\end{equation}

Again, the irreducible components of $d_{\mu\nu}$, $d_{(\mu\nu)}$ and $d_{[\mu\nu]}$, can be independently analyzed. The symmetric components generate
\begin{eqnarray}
\Delta H_{\rm{FW}}&=&- id_{00}\Sigma^i(1+\Phi)^{-2} \left[(1+\Phi)\partial_i + \frac{1}{2} (\partial_i \Phi)\right] +m d_{(0i)} \gamma^0\Sigma^i\nonumber\\
&&+\frac{\eta^{ij}}{2m}d_{(0i)}\gamma^0\Sigma^k \bigg[-(1+\Phi)\left(3-5\Phi\right) \partial_j \partial_k -\frac{1}{2}(1-15\Phi)((\partial_j \Phi) \partial_k+(\partial_k \Phi) \partial_j ) +\frac{3+5\Phi}{1+\Phi}(\partial_j \Phi)(\partial_k \Phi) \bigg]\nonumber\\
&& +i\frac{ \epsilon^{ijk }}{4m} d_{(0i)} \gamma^0 (1-\Phi) (\partial_j \Phi) \partial_k -i \eta^{jk} d_{(ij)} \Sigma^i\left[(
1+\Phi)\partial_k + \frac{1}{2}(\partial_k \Phi)\right],
\end{eqnarray}
and from the $d_{[\mu\nu]}$ components, the contribution to the nonrelativistic Hamiltonian is
\begin{eqnarray}
\Delta H_{\rm{FW}}&=&-m d_{[0i]} \gamma^0\Sigma^i+\frac{\eta^{ij}}{2m}d_{[0i]}\gamma^0\Sigma^k \big[-(1+\Phi)^2 \partial_j \partial_k  -\frac{3}{2}(1+\Phi)((\partial_j \Phi) \partial_k+(\partial_k \Phi) \partial_j )-(\partial_j \Phi)(\partial_k \Phi) \Big]\nonumber\\
&& -i\frac{ \epsilon^{ijk }}{4m} d_{[0i]} \gamma^0 (1-\Phi) (\partial_j \Phi) \partial_k  -i \eta^{jk} d_{[ij]} \Sigma^i\left[(
1+\Phi)\partial_k + \frac{1}{2}(\partial_k \Phi)\right].
\end{eqnarray}

The corresponding addition to the Hamiltonian coming from $e_\mu$ is
\begin{equation}
\Delta H_{\rm{FW}}=-m e_0 - \frac{i}{4m}{\epsilon_k}^{ij} e_0 \Sigma^k (1-3\Phi)(\partial_i \Phi) \partial_j  - \frac{1}{4m}\eta^{ij}e_0 (\partial_i \Phi)(\partial_j \Phi) +i\eta^{ij} e_i \gamma^0\left[ (1+\Phi)\partial_j+\frac{1}{2}(\partial_j \Phi)\right],
\end{equation}
and $f_\mu$ produces 
\begin{equation}
\Delta H_{\rm{FW}}=\frac{1}{2}f_0 \gamma^0 \Sigma^i (1+\Phi)^{-1}(\partial_i \Phi)+i\frac{\eta^{ij}f_i}{4m}\Sigma^k (1+\Phi)(1-3\Phi)\left[(\partial_j\Phi) \partial_k-(\partial_k\Phi) \partial_j\right].
 \end{equation}

The $g_{\mu\nu\rho}$ coefficient can be irreducibly decomposed \cite{g decomp paper} as
\begin{subequations}\label{gdecomp}
\begin{equation}
 g_{\mu\nu\rho}=g^{(A)}_\sigma{\varepsilon_{\mu\nu\rho}}^\sigma+\frac{2}{3}g^{(T)}_{[\mu} g_{\nu]\rho}+g^{(M)}_{\mu\nu\rho},
\end{equation}
where
\begin{eqnarray}
 g^{(A)}_\mu&=&\frac{1}{6}g_{\nu\rho\sigma}{\varepsilon_\mu}^{\nu\rho\sigma},\\
 g^{(T)}_\mu&=&g^{\rho\sigma}g_{\mu\rho\sigma},\\
 g^{(M)}_{\mu\nu\rho}&=&\frac{1}{3}\Big(2 g_{\mu\nu\rho}-g_{\rho\mu\nu}-g_{\nu\rho\mu} +g_{\mu\rho}g^{\alpha\beta}g_{\nu \alpha\beta} -g_{\nu\rho}g^{\alpha\beta}g_{\mu\alpha\beta}\Big)
\end{eqnarray}
\end{subequations}
are respectively known as the axial, trace and mixed-symmetry parts. The mixed-symmetry part satisfies $g^{(M)}_{\mu\nu\rho}{\varepsilon_\sigma}^{\mu\nu\rho}$ $=0$ and $g^{(M)}_{\mu\rho\sigma}g^{\rho\sigma}=0$, but it cannot be written in a simple manner in terms of $g_{\mu\nu\rho}$ and geometric tensors. Therefore, even though it is the natural separation from the point of view of the discussion presented in Appendix \ref{FieldRedef}, the separation in irreducible components is not practical for the calculation at hand. Instead, the effects of the components $g_{0i0}$, $g_{0ij}$, $g_{ij0}$, and $g_{ijk}$ are studied separately. In the $g_{0i0}$ case,
\begin{equation}
\Delta H_{\rm{FW}}= \frac{\eta^{ij}}{2}g_{0i0}\gamma^0(1+\Phi)^{-2}(\partial_j \Phi) - i{\epsilon_m}^{ij}g_{0i0}\gamma^0\Sigma^m(1+\Phi)^{-2}\left[(1+\Phi)\partial_j +\frac{1}{2}(\partial_j \Phi)\right].
\end{equation}
The presence of $g_{0ij}$ generates
\begin{eqnarray}
\Delta H_{\rm{FW}}&=&-i\frac{\eta^{ik}\eta^{jl}}{2m}g_{0ij}(1-3\Phi)(\partial_k \Phi) \partial_l\nonumber\\
&&+\frac{{\epsilon_m}^{il}\eta^{jk}}{2m}g_{0ij}\Sigma^m\Big[-2(1-\Phi^2) \partial_k \partial_l -(1-3\Phi)\left((\partial_k \Phi) \partial_l+(\partial_l \Phi) \partial_k\right) +\frac{1+2\Phi}{1+\Phi}(\partial_k \Phi)(\partial_l \Phi)\Big].
\end{eqnarray}
The terms produced by $g_{ij0}$ are
\begin{eqnarray}
\Delta H_{\rm{FW}}&=&-\frac{m}{2}{\epsilon_m}^{ij}g_{ij0}\Sigma^m-i \frac{\eta^{ik}\eta^{jl}}{8m}g_{ij0}(1+9\Phi)\left[(\partial_k \Phi) \partial_l - (\partial_l \Phi) \partial_k\right]\nonumber\\
&&-\frac{\eta^{kl}{\epsilon_m}^{ij}}{4m}g_{ij0}\Sigma^m \Big[(1+\Phi)^2 \partial_k \partial_l +3(1+\Phi) (\partial_k \Phi) \partial_l+(\partial_k \Phi)(\partial_l \Phi)\Big]\nonumber\\
&&+ \frac{ \eta^{ik}{\epsilon_m}^{jl}}{2m}g_{ij0}\Sigma^m\Big[-2(1-\Phi^2) \partial_k \partial_l-(1-3\Phi)((\partial_k \Phi) \partial_l +(\partial_l \Phi) \partial_k)+\frac{1+2\Phi}{1+\Phi}(\partial_k \Phi)(\partial_l \Phi)\Big]\nonumber\\
&&+\frac{\epsilon^{ijk}}{8m}g_{ij0}\Sigma^l \Big[2 (1+\Phi)^2 \partial_k \partial_l+3(1+\Phi) \left((\partial_k \Phi) \partial_l +(\partial_l \Phi) \partial_k\right)+(\partial_k \Phi)(\partial_l \Phi)\Big],
\end{eqnarray}
and the $g_{ijk}$ components produce
\begin{equation}
\Delta H_{\rm{FW}}= \frac{i}{2}\eta^{kl}{\epsilon_m}^{ij}g_{ijk}\gamma^0 \Sigma^m \left[(1+\Phi) \partial_l +\frac{1}{2}(\partial_l \Phi)\right].
\end{equation}
Finally, the contribution of $H_{\mu\nu}$ is
\begin{eqnarray}
\Delta H_{\rm{FW}}&=&-\frac{ \eta^{ij} }{4m}H_{0i}(1-3\Phi)(\partial_j\Phi) +i\frac{{\epsilon_k}^{ij}}{2m}H_{0i}\Sigma^k \left[2(1-\Phi^2) \partial_j +(1-3\Phi)(\partial_j\Phi)\right]\nonumber\\
&&+\frac{1}{2}H_{ij}{\epsilon_k}^{ij}\gamma^0 \Sigma^k(1+\Phi) +\frac{2\eta^{ik}{\epsilon_m}^{jl}+\eta^{kl}{\epsilon_m}^{ij}}{16m^2}H_{ij}\gamma^0 \Sigma^m\times(1+\Phi)(\partial_k\Phi)(\partial_l\Phi)\nonumber\\
&&+\frac{\epsilon^{ijk}}{16 m^2}H_{ij}\gamma^0 \Sigma^l \Big[4 (1+\Phi)^3 
\partial_k \partial_l  + 6 (1+\Phi)^2 \left((\partial_k\Phi) \partial_l + (\partial_l\Phi) \partial_k\right) +3 (1+\Phi)(\partial_k\Phi) (\partial_l\Phi) 
\Big].\label{contribution H}
\end{eqnarray}

To first order in $m_5$ and the SME coefficients, the nonrelativistic Hamiltonian $H_{\rm{FW}}$ is composed of the Lorentz-invariant Hamiltonian (\ref{H LI}) plus the corrections given in Eqs.~(\ref{contribution m5})-(\ref{contribution H}). Observe that each SME coefficient couples with spin and gravity in a particular way, and, in principle, every term in these expressions can be used to look for the coefficients' effects. However, before seeking these effects, it is convenient to analyze which terms in $H_{\rm{FW}}$ are not physical. This is done next. For simplicity, for the rest of the manuscript, all the equations are only valid to linear order in $\Phi$ and $\partial_i \Phi$. Also, the nonrelativistic Hamiltonian $H_{\rm{FW}}$ is written in terms of the momentum operator $p_i=-i \partial_i$, which is assumed to act on everything on its right.
 
\subsection{Physical terms}

As discussed in Appendix \ref{FieldRedef}, it can be shown that some combinations of SME coefficients cannot have physical meaning. This analysis is done by redefining the spinor field at the level of the action. Nevertheless, to this point all the SME coefficients have been preserved to provide an additional method to check the resulting Hamiltonian. The idea is that all the unphysical combinations of SME coefficients must cancel through unitary transformations.

Any unitary transformation $U= e^{iS}$ has to be even, to avoid coupling particles and antiparticles, and, to be compatible with the nonrelativistic approximation, it must have two or fewer powers of momentum. Under these conditions, the most general $S=S^\dagger$ is
\begin{eqnarray}
S &=& A + B \gamma^0 + C_i \Sigma^i + D_i\gamma^0 \Sigma^i+ \left(E^i + F^i \gamma^0 + G^i_j \Sigma^j + I^i_j \gamma^0 \Sigma^j\right)\frac{p_i}{m}\nonumber\\
&& +\left(J^{ij}+ K^{ij}\gamma^0+ L^{ij}_k \Sigma^k + M^{ij}_k \gamma^0 \Sigma^k \right)\frac{p_i}{m}\frac{p_j}{m},\label{S operator general}
\end{eqnarray}
where it is assumed that $A$, $B$, etc., are real, linear in the SME coefficients and may depend on $\Phi$. Unitary transformations in the Lorentz-invariant case are analyzed in Ref.~\cite{Obukhov}. To leading order in the SME coefficients, the result of this unitary transformation is
\begin{eqnarray}
H_{\rm{FW}}'&=& e^{iS}H_{\rm{FW}}e^{-iS}-i e^{iS}\partial_0 e^{-iS} \nonumber\\
&=&H_{\rm{FW}}+i [S, H_{\rm{FW,LI}}]-(\partial_0 S).
\end{eqnarray}

At this stage, a generic expression for the terms in $S$ is needed. At the corresponding order of approximation such expressions can be used to check that the first partial derivatives with respect to the spacetime coordinates are proportional to the SME coefficient contracted with $(\partial_i\Phi)$ and the second derivatives vanish. Therefore, $H_{\rm{FW}}'$ contains terms like $F^i(\partial_i\Phi)-(\partial_0 A)$, and
\begin{equation}
\left[-\eta^{ij} (\partial_i B)+(K^{ij}+K^{ji})(\partial_i\Phi)-(\partial_0E^j)\right] \frac{p_j}{m}.
\end{equation}
These terms can be chosen to cancel any real term in the Hamiltonian containing an SME coefficient, a factor of either $(\partial_i\Phi)$ or $(\partial_i\Phi) p_j$, and that is proportional to the identity matrix. Analogous derivations permit the conclusion that the same is valid for any even Dirac matrix. Therefore, any real term in $H_{\rm{FW}}$ proportional to an SME coefficient and either $(\partial_i\Phi)$ or $(\partial_i\Phi) p_j$ can be removed. Observe that the factor $-i(\partial_i\Phi)=p_i \Phi-\Phi p_i $ is not real, and thus it cannot be removed with these transformations.

After the unphysical terms are removed, the nonrelativistic Hamiltonian can be brought to the form
\begin{eqnarray}
H'_{\rm{FW}} &=&\left(m\gamma^0+\hat{a}_0-m\hat{c}_{00}\gamma^0-m \hat{e}_0\right)(1+\Phi)+\left(-b_k+m \hat{d}_{k0} \gamma^0+\frac{{\epsilon_k}^{lm}}{2}\left(-m\hat{g}_{lm0}+H_{lm}\gamma^0 \right) \right)\Sigma^k(1+\Phi) \nonumber\\
&&+ \frac{\eta^{il}}{2m}\left[a_l\gamma^0 -m(\hat{c}_{0l}+\hat{c}_{l0})-m e_l\gamma^0 \right]\left[(1+\Phi) p_i+ p_i (1+\Phi)\right] \nonumber\\
&& +\frac{1}{2m}\left[-\delta_k^i\hat{b}_0\gamma^0+m \delta_k^i\hat{d}_{00} +m\eta^{il} d_{kl} +{\epsilon_k}^{il}\left(-\frac{1}{2}\gamma^0 (\partial_l\Phi) -m \hat{g}_{0l0}\gamma^0+ \hat{H}_{0l} \right)-\frac{m}{2}\eta^{in}{\epsilon_k}^{lm}g_{lmn}\gamma^0\right]\nonumber\\
&&\times\Sigma^k \left[(1+\Phi) p_i+ p_i (1+\Phi)\right] \nonumber\\
&& +\frac{1}{2m}\left(\eta^{ij}\gamma^0 -\eta^{ij}\hat{c}_{00}\gamma^0 -\eta^{il}\eta^{jm}(c_{lm}+c_{ml}) \gamma^0\right)p_{(i} (1+\Phi) p_{j)} \nonumber\\
&&+\frac{1}{2m^2}\left[\eta^{ij} b_k - \eta^{il}\delta_k^j b_l -m \eta^{il}\delta^j_k\hat{d}_{l0}\gamma^0 + \frac{m}{2}{\epsilon_k}^{lm}\eta^{ij}\hat{g}_{lm0} - \frac{ \epsilon^{ilm} \delta_k^j }{2}\left(m\hat{g}_{lm0}+H_{lm}\gamma^0\right)\right]\Sigma^k p_{(i}(1+3\Phi) p_{j)} \nonumber\\
&& +\frac{1}{m} \left[\eta^{il}\delta^j_k(\hat{d}_{0l}+\hat{d}_{l0})\gamma^0 + {\epsilon_k}^{il}\eta^{jm}(\hat{g}_{lm0}-\hat{g}_{0lm})\right]\Sigma^k p_{(i} (1+\Phi) p_{j)},\label{Hamiltonian 4x4}
\end{eqnarray}
where the coefficients with a caret get a factor $1-n\Phi$, $n$ being the number of indices that are zero. For example,
\begin{eqnarray*}
 \hat{a}_0&=&(1-\Phi)a_0,\quad \hat{c}_{k0}=(1-\Phi)c_{k0},\\
 \hat{d}_{00}&=& (1-2\Phi)d_{00},\quad \hat{H}_{ij}=H_{ij}.
\end{eqnarray*}
Recall that Eq.~(\ref{Hamiltonian 4x4}) is only valid to linear order in $\Phi$. Also, observe that in Eq.~(\ref{Hamiltonian 4x4}) all the SME coefficients have a caret.

One advantage of using the caret notation is that the linear combinations of SME coefficients appearing in the Hamiltonian (\ref{Hamiltonian 4x4}) coincide with those of Ref.~\cite{KosteleckyLane} where gravity is disregarded (up to signs from using different metric signatures). Therefore, the result of this work is that the effects of a uniform gravitational field can be introduced through redshift factors in the rest energy term, the momentum, and the SME coefficients. It is easy to verify the SME coefficients get a factor $(1+\Phi)^{-1}$ for each zero index. However, it is not straightforward to understand the couplings of the gravitational potential and the momentum. In particular, it would be interesting to understand the separation of the terms quadratic in momentum that contain $\Sigma^k$ into those where the gravitational factor is $1+3\Phi$ and those where the factor is $1+\Phi$. Clearly, to grasp these issues, it is necessary to keep higher order terms in $\Phi$, which lies outside the scope of this paper.

\subsection{Consistency checks}

In this section, some limits of the Hamiltonian (\ref{Hamiltonian 4x4}) and consistency checks are considered. First, the Hamiltonian (\ref{Hamiltonian 4x4}) coincides with the one of Hehl and Ni \cite{Hehl90} in the limit where all the SME coefficients are set to zero (provided that $\sigma^i$ in Ref.~\cite{Hehl90} is identified with $\gamma^0 \Sigma^i$). In addition, in the limit where $\Phi=0$, the Hamiltonian (\ref{Hamiltonian 4x4}) agrees with the nonrelativistic Hamiltonian of Refs.~\cite{KosteleckyLane,KosteleckyLane2}. Moreover, the calculation of spin-independent SME coefficients in a general gravitational field \cite{KosteleckyTasson} is also used to compare, and, where there is overlap, the nonrelativistic Hamiltonian of Ref.~\cite{KosteleckyTasson} coincides with the Hamiltonian (\ref{Hamiltonian 4x4}).

Note that $\gamma^0$ only appears in the Hamiltonian (\ref{Hamiltonian 4x4}) in terms having no SME coefficients and an even number of momentum (and derivatives) or in terms having a coefficient with an odd (even) number of indices and odd (even) powers of momentum. This is closely related to the result \cite{ColladayKostelecky} that coefficients with an odd (even) number of indices are CPT odd (even). Also, causality, the loss of unitarity, and other related issues should not present additional complication than in the nongravitational SME \cite{flat spacetime issues} because, after all, in this work, spacetime is flat.

In the representation of the Dirac matrices that is used, the particle Hamiltonian corresponds to the $2\times2$ upper-left block of Eq.~(\ref{Hamiltonian 4x4}). In practice, this Hamiltonian is obtained by replacing the $4\times 4$ identity matrix and $\gamma^0$ with the $2\times 2 $ identity matrix and $\Sigma^k$ with the Pauli matrices $\sigma^k$. Since the coefficient structure is compatible with that of the gravity-free case, the antiparticle Hamiltonian can be generated from the particle Hamiltonian with the same replacements discussed in Ref.~\cite{KosteleckyLane}.

As mentioned in Appendix \ref{FieldRedef}, the observable combinations of coefficients can be obtained from those observable combinations in the gravity-free case by writing carets over the coefficients. This automatically guarantees that the Hamiltonian (\ref{Hamiltonian 4x4}) is compatible with the restrictions coming from the freedom to redefine the fields at the level of the action.

Note that the Hamiltonian (\ref{Hamiltonian 4x4}) is Hermitian with respect to the standard inner product of nonrelativistic quantum mechanics. To check this, it has to be considered that $p_i$ acts on the SME coefficients in the self-adjoint Hamiltonian. However, the assumption that the SME coefficients have vanishing covariant derivatives can be used to verify that the momentum operator acting on any coefficient with a caret vanishes.

Moreover, the Hamiltonian (\ref{Hamiltonian 4x4}) is not explicitly invariant under the gauge transformation associated with $\Phi$. Adding a constant to $\Phi$ amounts to changing the height where $\Phi=0$, which, in turn, is associated with the position of the accelerated observer. Thus, to check that the physics is invariant, it is necessary to consider that, under such transformation, the height of the observer changes and time at this new height gets redshifted. If this effect is considered, it is possible to verify that the physics is invariant under such transformation. In the next section, some comments on the experimental implications of the Hamiltonian (\ref{Hamiltonian 4x4}) are given.

\section{Experimental consequences}\label{ExpConseq}

To date, there is no compelling experimental evidence of unconventional effects associated with gravity, even when spin-polarized matter is used (see, e.g., Ref.~\cite{Adelberger}). Nevertheless, empirical tests where gravity plays an important role, including those associated with Lorentz violation, may become relevant as they could uncover new physics. In what follows, some proposals to empirically look for the effects of the Hamiltonian (\ref{Hamiltonian 4x4}) are briefly discussed.

Atomic interferometers are a class of experiments that are sensitive to the gravitational field \cite{Muller} and which have been used to set bounds on SME coefficients \cite{Hohensee11}. In fact, these experiments offer one of the most compelling techniques to test for spin-insensitive coefficients \cite{Hohensee12}. One interesting possibility is to modify these experiments and make them sensitive to spin. However, it seems unlikely that these experiments could compete with the experiments that have set bounds on spin-dependent coefficients \cite{boundsb}.

In laboratory experiments, coefficients associated with higher orders of momentum are typically more difficult to constrain. The terms in the Hamiltonian (\ref{Hamiltonian 4x4}) with $n$ powers of momentum have a remaining piece that has $n-1$ momentum operators since one of the $p_i$ can act on $\Phi$. In principle, this could be used to constrain SME coefficients associated with certain powers of momentum with experiments sensitive to fewer momentum powers. Nevertheless, this seems unfeasible since the gradient of the gravitational field on Earth is incredibly small. In addition, astrophysical observations \cite{astrophys} can be used to test the regime where the higher order in momentum terms dominate.

It is noteworthy that the first line in Eq.~(\ref{Hamiltonian 4x4}) behaves like a spin-dependent mass. Therefore, experiments testing the universality of freefall \cite{UFF} could become sensitive to the parameters in those terms, particularly if the experiments are modified to become sensitive to spin polarization.

An intriguing consequence of the Newtonian gravitational potential is that, in any experiment done on Earth, the bounds are actually set on linear combinations of SME coefficients with carets. Namely, each coefficient in these linear combinations has a different power of the factor $1+\Phi$. Since $\Phi$ depends on the altitude, by doing experiments at different heights, it should be possible to translate the constraints from linear combinations into individual bounds. One of the main advantages is that, to disentangle the bounds, there is no need to modify the experiments. In addition, this can be used for all the SME coefficients, including those that are spin sensitive, since $\Phi$ affects all the SME coefficients in the same way.

That gravity allows one to translate bounds on combinations of SME coefficients into individual constraints resembles the observation \cite{Altschul} that different energies can also be used to disentangle bounds. This result can be traced to the fact that the Lorentz factor couples differently with each SME coefficient \cite{KosteleckyLane}. Here the gravitational field is introduced through a uniformly accelerated observer, which, in turn, is related to an inertial observer through a series of boosts, thus, it is not surprising that a similar result is found.

As mentioned above, some SME coefficients acquire a sign when changing from particle to antiparticle. This could be also used to separate bounds \cite{Tasson12}. Of course, experiments with antimatter are daunting. Nevertheless, freefall tests with antimatter will be performed in the near future \cite{antihydrogen}. It is most likely that, to disentangle all SME bounds, every possible mechanism would be needed, and gravity could play a highly unanticipated role in the search for Lorentz violation.

\section{Conclusions}\label{Conclusions}

The nonrelativistic Hamiltonian for the minimal matter SME sector in the presence of gravity has been derived, including, for the first time, spin-sensitive SME coefficients. The basic ideas used throughout the calculation are that a uniform Newtonian gravitational field is an accurate description of gravity for most laboratory experiments, and that flat spacetime as described by a uniformly accelerated observer models this gravitational environment. The spinors are taken as test particles, and in this approximation it is consistent to neglect coefficient fluctuations, which simplifies the calculations dramatically. 

One of the consequences of this analysis is that, since gravity couples to each SME coefficient in a different way, by doing experiments at different altitudes, it could be possible to separate the bounds from constraints on linear combinations of coefficients into individual constraints. It should be mentioned that disentangling these bounds would exhaust the possibility that some SME coefficients are nonzero but, for some reason, the effect of these coefficients cancels. Also, if there is some day a positive signal for Lorentz violation, when trying to come out with a fundamental explanation, it would be necessary to know which coefficients are responsible for such a signal.

At the classical level, the effects of explicit SME coefficients in the presence of gravity have been incorporated in the framework of pseudo-Riemann--Finsler geometry \cite{Finsler1,Finsler2}. Since there is a considerable overlap in the assumptions of this work and those taken in the pseudo-Riemann--Finsler treatment, it is tempting to understand the connections between these approaches. Other extensions to this work are, for instance, generalizing the method to any static background spacetime. In this case, it would still be possible to use the test-particle approximation and consistently neglect the coefficient fluctuations; however, the assumption that the coefficients have vanishing covariant derivatives would only be valid locally. Also, it seems interesting to generalize the method to other SME sectors, including the neutrino \cite{neutrinos} and nonminimal sectors \cite{nonminimal}, where operators of arbitrary dimension are considered.

\subsection*{Acknowledgments}
I thank Alan Kosteleck\'y and Jay Tasson for many encouraging and clarifying discussions and Ralf Lehnert for his suggestions on the manuscript. This work was supported in part by U.S. DOE Grant No. DE-FG02-13ER42002 and by the Indiana University Center for Spacetime Symmetries.

\appendix

\section{Notation and conventions}\label{Notation}

The notation and conventions used throughout the paper, which follow closely Ref.~\cite{Kostelecky04}, are explained here. Greek letters are used as spacetime indices and the spacetime metric, $g_{\mu\nu}$, has signature $(-+++)$. As is customary, the metric and its inverse, $g^{\mu\nu}$, are used to lower and raise spacetime indices. The covariant derivative and the spacetime volume $4$-form associated with $g_{\mu\nu}$ are respectively denoted by $\nabla_\mu$ and $\varepsilon_{\mu\nu\rho\sigma}$. Note that $\nabla_\kappa \varepsilon_{\mu\nu\rho\sigma}=0$. As usual, $\partial_\mu$ is used to represent the derivative with respect to the coordinates.

The tetrad is a set of vectors $\{e^\mu_a\}$ satisfying $g_{\mu\nu}e^\mu_a e^\nu_b=\eta_{ab}$, where $\eta_{ab}=\rm{diag}(-1,1,1,1)$. Note that the Latin indices from the beginning of the alphabet ($a,b,c,d,e,f$) can be thought of as tangent-space indices, making them appropriate for the Dirac matrices. These indices run from $0$ to $3$, $0$ being the temporal coordinate. Latin indices of the middle of the alphabet ($i,j\ldots,n$) refer only to the spatial components and run from $1$ to $3$. Also, $\eta_{ab}$ and its inverse, $\eta^{ab}$, can be used to lower and raise tangent-space indices. Any $n$ indices inside parentheses (brackets) denote symmetrization (antisymmetrization) with a factor $1/(n!)$.

Given that the tetrad is an orthonormal (and right-handed) basis, the components of the spacetime volume form in this basis,
\begin{equation} \label{volume form vs totally antisym}
 \epsilon_{abcd}=\varepsilon_{\mu\nu\rho\sigma}e^\mu_a e^\nu_b e^\rho_c e^\sigma_d,
\end{equation}
may be regarded as the totally antisymmetric tensor with the convention that $\epsilon_{0123}=1$. The components $\epsilon_{0ijk}$ are written as $\epsilon_{ijk}$.

The spin connection is defined as 
\begin{equation}\label{spin connection}
\omega_{\mu ab}=g_{\rho\sigma} e^\rho_a \nabla_\mu e^\sigma_b=-\omega_{\mu ba}.
\end{equation}
It is possible to show that $\nabla_\mu e^\mu_a= \eta^{cd}e^\mu_c \omega_{\mu d a}$. Thus, $\nabla_\mu$ is not the same derivative operator as the $D_\mu$ used in Ref.~\cite{Kostelecky04}, which annihilates the tetrad. This discrepancy is compensated by the definition of the spin connection (\ref{spin connection}). In terms of the tetrad, the spacetime volume element is denoted by $e$.

The Dirac matrices $\gamma^a$ are taken to satisfy
\begin{equation}
\{\gamma^a,\gamma^b\}=-2\eta^{ab}.
\end{equation}
The matrices $\sigma^{ab}$ and $\gamma_5$ are defined by
\begin{eqnarray}
\sigma^{ab}&=&\frac{i}{2}[\gamma^a,\gamma^b],\\ \gamma_5&=&i\gamma^0\gamma^1\gamma^2\gamma^3,
\end{eqnarray}
and the spin matrix is
\begin{equation}
 \Sigma^i=\gamma_5 \gamma^0 \gamma^i=\frac{1}{2}{\epsilon_{kl}}^i\sigma^{kl}.
\end{equation}
Throughout the paper, the Pauli--Dirac representation for the Dirac matrices is assumed. It is advantageous to use the basis of Dirac matrices $1$, $\gamma^0$, $\gamma^i$, $\gamma_5$, $\gamma^0\gamma^i$, $\gamma^0\gamma_5$, $\Sigma^i$, $\gamma^0 \Sigma^i$, where $1$, $\gamma^0$, $\Sigma^i$, $\gamma^0 \Sigma^i$ are even (and Hermitian) and the rest are odd. Finally, the covariant derivative acting on spinors takes the form
\begin{equation}
\nabla_\mu \psi=\partial_\mu \psi+\frac{i}{4}\omega_{\mu ab}\sigma^{ab}\psi,\qquad
\nabla_\mu \bar{\psi}=\partial_\mu \bar{\psi} -\frac{i}{4}\omega_{\mu ab}\bar{\psi}\sigma^{ab}.
\end{equation}

\section{Field redefinitions}\label{FieldRedef}

The goal of this Appendix is to identify the combinations of SME coefficients which can have physical consequences since they cannot be canceled with field redefinitions. Note that the method presented here is slightly different from the one discussed by other authors \cite{ColladayKostelecky,Kostelecky04,KosteleckyTasson09,KosteleckyTasson,Finsler1,field redef bibliog}, where spurious SME coefficients are generated through field redefinitions. The analysis is first done in a general spacetime, and the specific field redefinitions relevant for this paper are then presented. The starting point is the action (\ref{action}). As is discussed in Ref.~\cite{Kostelecky04}, the spinor field can redefined as
\begin{equation}
 \psi = \left(1+V\right)\chi,
\end{equation}
where $\chi$ is a spinor field and
\begin{eqnarray}
V &=& \left(v^{(1)}_\mu +i v^{(2)}_\mu\right) e^\mu_a \gamma^a+ \left(v^{(3)}_\mu +i v^{(4)}_\mu\right) e^\mu_a \gamma_5\gamma^a +\left(v^{(5)}_{\mu\nu} +i v^{(6)}_{\mu\nu}\right) e^\mu_a e^\nu_b \sigma^{ab},
\end{eqnarray}
with $v^{(5)}_{\mu\nu}=-v^{(5)}_{\nu\mu}$ and $v^{(6)}_{\mu\nu}=-v^{(6)}_{\nu\mu}$. If $V$ is linear in the SME coefficients, the action for $\chi$, to first order in the Lorentz-violating coefficients, takes the form
\begin{eqnarray}
S&=& \int d^4 x e \Big[\frac{i}{2} e^\mu_a \bar{\chi}(\Gamma^a+\Delta \Gamma^a) \nabla_\mu\chi -\frac{i}{2} e^\mu_a (\nabla_\mu\bar{\chi})(\Gamma^a+\Delta \Gamma^a) \chi-\bar{\chi}(M+\Delta M)\chi\Big],\label{action redefined}
\end{eqnarray}
where
\begin{eqnarray}
\Delta\Gamma^a&=&\gamma^a V+V^* \gamma^a,\nonumber \\
&=&- 4v^{(6)}_{\mu\nu} \eta^{ac}e^\mu_b e^\nu_c \gamma^b - 2v^{(5)}_{\mu\nu} {\epsilon^{acd}}_b e^\mu_c e^\nu_d \gamma_5 \gamma^b-2 v^{(1)}_\mu \eta^{ab} e^\mu_b + 2iv^{(4)}_\mu \eta^{ab} e^\mu_b \gamma_5\nonumber\\
&&-\Big( 2 v^{(2)}_\mu e^\mu_{[c} \delta^a_{d]}+ v^{(3)}_\mu{\epsilon^{ab}}_{cd} e^\mu_b\Big)\sigma^{cd},
\end{eqnarray}
and
\begin{eqnarray}
\Delta M&=&m( V+ V^*) +\frac{i}{2}e^\mu_a \left(-\gamma^a(\partial_\mu V)+(\partial_\mu V^*) \gamma^a\right) +\frac{1}{8}e^\mu_a \omega_{\mu cd}\left( [V^*,\sigma^{cd}]\gamma^a-\gamma^a[V,\sigma^{cd}]\right)\nonumber\\
&=&-\nabla_\mu {v^{(2)}}^\mu-i\nabla_\mu {v^{(3)}}^\mu \gamma_5  +\left(2mv^{(1)}_\mu e^\mu_a-2 (\nabla_\mu v^{(5)}_{\nu\rho})g^{\mu\nu} e^\rho_a \right)\gamma^a +\left( 2m v^{(3)}_\mu e^\mu_a -{\epsilon^{bcd}}_a e^\mu_b e^\nu_c e^\rho_d \nabla_\mu v^{(6)}_{\nu\rho} \right)\gamma_5\gamma^a\nonumber\\
&&+\Big[-(\nabla_{[\mu} v^{(1)}_{\nu]}) e^\mu_a e^\nu_b -\frac{1}{2}(\nabla_\mu v^{(4)}_\nu){\epsilon^{cd}}_{ab} e^\mu_c e^\nu_d +2 mv^{(5)}_{\mu\nu} e^\mu_a e^\nu_b\Big] \sigma^{ab}.
\end{eqnarray}

The effective coefficients are defined as those in the action (\ref{action redefined}), concretely
\begin{eqnarray}
\Gamma^a + \Delta \Gamma^a&=&\gamma^a-c^{\rm eff}_{\mu\nu}\eta^{ac}e^\mu_b e^\nu_c \gamma^b-d^{\rm eff}_{\mu\nu}\eta^{ac}e^\mu_b e^\nu_c \gamma_5 \gamma^b  - e^{\rm eff}_\mu 
\eta^{ac}e^\mu_c-i f^{\rm eff}_\mu \eta^{ac}e^\mu_c \gamma_5 -\frac{1}{2}g^{\rm eff}_{\mu\nu\rho}\eta^{ab} e^\mu_c e^\nu_d e^\rho_b \sigma^{cd},\\
M+\Delta M &=&m^{\rm eff}+i m^{\rm eff}_5 \gamma_5+a^{\rm eff}_\mu e^\mu_a \gamma^a +b^{\rm eff}_\mu e^\mu_a \gamma_5 \gamma^a+\frac{1}{2}H^{\rm eff}_{\mu\nu} e^\mu_a e^\nu_b \sigma^{ab}.
\end{eqnarray}
The idea is to choose $V$ in such a way that most components of the effective coefficients cancel. However, to avoid generating spacetime-dependent $m^{\rm eff}$ and $m^{\rm eff}_5$, it is necessary to take
\begin{equation}\label{nabla v}
\nabla_\mu {v^{(2)}}^\mu=\nabla_\mu {v^{(3)}}^\mu=0.
\end{equation}
Incidentally, spacetime-dependent masses have been considered as alternative explanations for astrophysical observations \cite{Landau} and some neutrino anomalies \cite{AceroBonder}. In addition to Eq.~(\ref{nabla v}), it is convenient to set
\begin{equation}
v^{(1)}_\mu=-\frac{1}{2}e_\mu,\quad
v^{(4)}_\mu=\frac{1}{2}f_\mu,\quad
v^{(5)}_{\mu\nu}=-\frac{1}{8}{\varepsilon_{\mu\nu}}^{\rho\sigma} d_{\rho\sigma},\quad
v^{(6)}_{\mu\nu}=-\frac{1}{4}c_{[\mu\nu]}.
\end{equation}
With these choices the effective coefficients become
\begin{subequations}\label{eff coef general}
\begin{eqnarray}
m^{\rm eff}&=&m,\\
m_5^{\rm eff}&=&m_5,\\
a_\mu^{\rm eff}&=&a_\mu - m e_\mu - \frac{1}{4}{\varepsilon_\mu}^{\nu\rho\sigma}\nabla_\nu d_{\rho\sigma} , \\
b_\mu^{\rm eff} &=&b_\mu+2m v^{(3)}_\mu-\frac{1}{4}{\varepsilon_\mu}^{\nu\rho\sigma} \nabla_\nu c_{\rho\sigma},\\
c^{\rm eff}_{\mu\nu}&=&c_{(\mu\nu)},\\
d^{\rm eff}_{\mu\nu} &=&d_{(\mu\nu)},\\
e^{\rm eff}_\mu&=&0,\\
f^{\rm eff}_\mu&=&0,\\
g^{\rm eff}_{\mu\nu\rho} &=& g_{\mu\nu\rho} + 4 v^{(2)}_{[\mu} g_{\nu]\rho} + 2 {\varepsilon_{\mu\nu\rho}}^\sigma v^{(3)}_\sigma,\\
H^{\rm eff}_{\mu\nu}&=&H_{\mu\nu}-\frac{m}{2} {\varepsilon_{\mu\nu}}^{\rho\sigma} d_{\rho\sigma} + \nabla_{[\mu} e_{\nu]} -\frac{1}{2}{\varepsilon_{\mu\nu}}^{\rho\sigma} \nabla_\rho f_\sigma .
\end{eqnarray}
\end{subequations}

In a general case, it is suitable to set $v^{(2)}_\mu =v^{(3)}_\mu =0$, which leads to results that are in agreement with Ref.~\cite{Kostelecky04}. However, in this work, the SME coefficients have vanishing covariant derivatives, and $V$ can be chosen in such way that parts of $g_{\mu\nu\rho}$ cancel while, at the same time, Eq.~(\ref{nabla v}) is satisfied. In particular, the selection
\begin{equation}
v^{(2)}_\mu =-\frac{1}{6}g^{\rho\sigma}g_{\mu\rho\sigma},\qquad
v^{(3)}_\mu =-\frac{1}{12}{\varepsilon_\mu}^{\nu\rho\sigma}g_{\nu\rho\sigma},
\end{equation}
can be used to cancel the axial and trace components of $g_{\mu\nu\rho}$, defined in Eqs.~(\ref{gdecomp}). With these elections, $m^{\rm eff}$, $m_5^{\rm eff}$, $c^{\rm eff}_{\mu\nu}$, $d^{\rm eff}_{\mu\nu}$, $e^{\rm eff}_\mu$, and $f^{\rm eff}_\mu$ are given as in Eqs.~(\ref{eff coef general}), and the remaining effective coefficients are
\begin{subequations}
\begin{eqnarray}
a_\mu^{\rm eff}&=&a_\mu -m e_\mu , \\
b_\mu^{\rm eff} &=&b_\mu - \frac{m}{6}{\varepsilon_\mu}^{\nu\rho\sigma}g_{\nu\rho\sigma} ,\\
g^{\rm eff}_{\mu\nu\rho} &=&g^{(M)}_{\mu\nu\rho},\\
H^{\rm eff}_{\mu\nu}&=&H_{\mu\nu} -\frac{m}{2} {\varepsilon_{\mu\nu}}^{\rho\sigma} d_{\rho\sigma} ,
\end{eqnarray}
\end{subequations}
where $g^{(M)}_{\mu\nu\rho}$ is defined in Eqs.~(\ref{gdecomp}).

Note that the spacetime metric and volume form enter in the effective coefficients. This suggests that, by doing experiments in different gravitational environments, it may be possible to separate the effects of SME coefficients that combine into an effective coefficient. In the particular case of the metric (\ref{metric}), it can be shown that, for all the effective coefficients, the modification due to gravity is that all the coefficients have to be replaced by those having a caret as introduced in Sec. \ref{ExpConseq}. As an example, the components of $H^{\rm eff}_{\mu\nu}$ are derived. Using Eq.~(\ref{volume form vs totally antisym}), it is possible to check that
\begin{equation}
{\varepsilon_{\mu\nu}}^{\rho\sigma} ={\epsilon_{ab}}^{cd} e^\alpha_e e^\beta_f e^\rho_c e^\sigma_d g_{\mu\alpha}g_{\nu\beta} \eta^{ae} \eta^{bf}.
\end{equation}
Taking this into the account, it can be verified that
\begin{equation}
H^{\rm eff}_{0i}=H_{0i} -\frac{m}{2}(1+\Phi) {\epsilon_i}^{kl} d_{kl},\qquad
H^{\rm eff}_{ij}=H_{ij} +m {\epsilon_{ij}}^k(1+\Phi)^{-1} d_{[0k]},
\end{equation}
which, in turn, implies
\begin{equation}
\hat{H}^{\rm eff}_{0i} = \hat{H}_{0i} -\frac{m}{2} {\epsilon_i}^{kl} d_{kl},\qquad
H^{\rm eff}_{ij} = H_{ij} +m {\epsilon_{ij}}^k \hat{d}_{[0k]} .
\end{equation}
From these expressions it is possible to verify that the components of $\hat{H}^{\rm eff}_{\mu\nu}$ have the same functional form in terms of $\hat{H}_{\mu\nu}$ and $\hat{d}_{\mu\nu}$ as the corresponding effective coefficients have in terms of $H_{\mu\nu}$ and $d_{\mu\nu}$ when gravity is disregarded. Analogous conclusions can be reached for all the effective coefficients by doing the corresponding calculations.

\end{document}